\documentclass{agujournal}

\journalname{Geophysical Research Letters}
\usepackage[dvipdfmx]{}
\usepackage{url}

\begin{document}

\title{Termination of Electron Acceleration in Thundercloud\\ by Intra/Inter-cloud Discharge}

\authors{Y. Wada\affil{1, 2}, G. Bowers\affil{3}, T. Enoto\affil{4}, M. Kamogawa\affil{5}, Y. Nakamura\affil{6}, T. Morimoto\affil{7}, \\
D. M. Smith\affil{8}, Y. Furuta\affil{1}, K. Nakazawa\affil{9}, T. Yuasa\affil{10}, A. Matsuki\affil{11}, M. Kubo\affil{12}, \\
T. Tamagawa\affil{2}, K. Makishima\affil{13} and H. Tsuchiya\affil{14}}

\affiliation{1}{Department of Physics, Graduate School of Science, The University of Tokyo, Tokyo, Japan}
\affiliation{2}{High Energy Astrophysics Laboratory, Nishina Center, RIKEN, Saitama, Japan}
\affiliation{3}{Space and Remote Sensing Group, Los Alamos National Laboratory, Los Alamos, NM, USA}
\affiliation{4}{The Hakubi Center for Advanced Research and Department of Astronomy, Kyoto University, Kyoto, Japan}
\affiliation{5}{Department of Physics, Tokyo Gakugei Univeritym Tokyo, Japan}
\affiliation{6}{Department of Electric Engineering, Kobe City College of Technology, Hyogo, Japan}
\affiliation{7}{Department of Electric and Electronic Engineering, Kindai University, Osaka, Japan}
\affiliation{8}{Department of Physics, University of California Santa Cruz, CA, USA}
\affiliation{9}{Kobayashi-Maskawa Institute for the Origin of Particles and the Universe, Nagoya University, Nagoya, Japan}
\affiliation{10}{55 Devonshire Road, Singapore 239855, Singapore}
\affiliation{11}{Institute of Nature and Environmental Technology, Kanazawa University, Ishikawa, Japan}
\affiliation{12}{College of Science and Engineering, Kanazawa University, Ishikawa, Japan}
\affiliation{13}{MAXI Team, RIKEN, Saitama, Japan}
\affiliation{14}{Nuclear Science and Engineering Center, Japan Atomic Energy Agency, Ibaraki, Japan}

\correspondingauthor{Yuuki Wada}{wada@juno.phys.s.u-tokyo.ac.jp}

\begin{keypoints}
\item A gamma-ray glow and its termination with a lightning discharge was observed in a Japanese winter thunderstorm. 
\item The glow was terminated by leader development of a horizontally-long 
    intra/inter-cloud discharge passing nearby overhead.
\item The intra/inter-cloud discharge was not triggered by the glow 
    in the present case because it started far from the gamma-ray glow site.
\end{keypoints}

\begin{abstract}
An on-ground observation program for high energy atmospheric phenomena in winter thunderstorms along Japan Sea
    has been performed via measurements of gamma-ray radiation, atmospheric electric field and low-frequency radio band.
    On February 11, 2017, the radiation detectors recorded gamma-ray emission lasting for 75 sec.
    The gamma-ray spectrum extended up to 20~MeV and was reproduced 
    by a cutoff power-law model with a photon index of $1.36^{+0.03}_{-0.04}$, being consistent with a Bremsstrahlung radiation from a thundercloud
    (as known as a gamma-ray glow and a thunderstorm ground enhancement).
    Then the gamma-ray glow was abruptly terminated with a nearby lightning discharge.
    The low-frequency radio monitors, installed $\sim$50~km away from the gamma-ray observation site recorded leader development of an intra/inter-cloud discharge 
    spreading over $\sim$60 km area with a $\sim$300 ms duration.
    The timing of the gamma-ray termination coincided with the moment 
    when the leader development of the intra/inter-cloud discharge 
    passed 0.7 km horizontally away from the radiation monitors.
    The intra/inter-cloud discharge started $\sim$15~km away from the gamma-ray observation site.
    Therefore, the glow was terminated by the leader development, 
    while it did not trigger the lightning discharge in the present case.
\end{abstract}

\section{\label{sec:intro}Introduction}
Gamma-ray glows are long-duration gamma-ray emissions 
    with energy reaching up to several tens of MeV associated with thunderstorm activities. 
    They have been observed inside thunderclouds by airplane and balloon experiments \citep{McCarthy_1985,Eack_1996,Kelley_2015}, 
    under thunderclouds by high-mountain experiments
    \citep{Brunetti_2000,Torii_2009,Tsuchiya_2009,Tsuchiya_2012,Chilingarian_2010,Chilingarian_2011,Chilingarian_2016},
    as well as by sea-level measurements \citep{Torii_2002,Torii_2011,Tsuchiya_2007,Tsuchiya_2011,Kuroda_2016}.
    Gamma-ray grows are also referred as long bursts \citep{Torii_2011}, 
    and thunderstorm ground enhancements (TGEs: \citet{Chilingarian_2011}) when detected by on-ground measurements.
    The phenomena typically last for several minutes, and are not generally accompanied with lightning. 
    Gamma rays are thought to be produced by the Relativistic Runaway Electron Avalanche process 
    (RREAs; \citet{Gurevich_1992,Dwyer_2003b,Kelley_2015}): energetic electrons seeded by e.g. cosmic rays 
    are accelerated by strong electric fields in thunderclouds, 
    and produce secondary electrons by ionizing ambient atmosphere. 
    The accelerated and multiplied electrons produce Bremsstrahlung photons in the atmosphere.
    We consider the observations that have been called ``gamma-ray glows'', ``long bursts'' and ``TGEs'' in the literature 
    to all be cases of RREAs and their daughter products (including gamma-rays and, when detectable, neutrons) 
    taking place in a strong thundercloud field.

There are reports of gamma-ray glows and TGEs abruptly terminated by lightning discharges
    \citep{McCarthy_1985,Eack_1996,Alexeenko_2002,Tsuchiya_2013,Kelley_2015,Chilingarian_2015,Chilingarian_2017}.
    In the past studies, temporal relation between such a glow and lightning \citep{Tsuchiya_2013} and 
    types of lightning to terminate glows \citep{Chilingarian_2017} have been discussed. 
    Gamma-ray glows, as evidence of stable electric-field particle acceleration in thunderclouds, 
    give us a few intriguing questions to be revealed: (i) where is the electron-acceleration region located? 
    (ii) which structure of thunderclouds corresponds to the acceleration region? 
    (iii) how does the acceleration region emerge, grow and disappear? 
    The sudden extinction of the acceleration region with lightning 
    can provide a hint for the location of the acceleration region because location of discharges can be well monitored with radio bands. 
    
We focus on high energy phenomena in winter thunderstorms along the coast of Japan Sea 
    which has unique characteristics such as low cloud bases and large discharge currents.
    We have continued gamma-ray radiation and atmospheric electric-field (AEF) measurements in this area.
    In the present paper, we report the first simultaneous detection of a gamma-ray glow termination
    with a lightning mapping observation in the low-frequency band.

\section{\label{sec:observation}Observation}
Our observation site is at Kanazawa University Noto School
    (37$^{\circ}$27'04''~N, 137$^{\circ}$21'32''~E), 
    located in the northern edge of Noto Peninsula in Japan. 
    We operated two independent gamma-ray detectors on the roof of the building with a 40-m separation between them.
    Detector A deployed by the GROWTH (Gamma-Ray Observation of Winter Thundercloud) collaboration \citep{Enoto_2017} 
    has a Bi$_{4}$Ge$_{3}$O$_{12}$ (BGO) scintillation crystal ($\phi$7.62~cm $\times$ 7.62~cm) coupled with a photo-multiplier tube (PMT), observing 0.2--7.0~MeV photons. 
    Detector B deployed by the GODOT (Gamma-ray Observations During Overhead Thunderstorms) 
    collaboration \citep{Bowers_2017} has a NaI scintillation crystal ($\phi$12.7~cm $\times$ 12.7~cm) 
    coupled with a PMT, observing 0.3--20.0~MeV photons. 
    Detector B also has small ($\phi$2.54~cm $\times$ 2.54~cm) 
    and large ($\phi$12.7~cm $\times$ 12.7~cm) plastic scintillators for neutron and gamma-ray detection at very high count rates, 
    and a blank phototube which is not coupled with any scintillation crystals for noise monitoring.
    Because most of signals seem to originate from gamma rays, we concentrate on results 
    from the gamma-ray sensitive NaI and BGO scintillators in the present analysis.

Both detectors A and B record energy deposits and arrival time of each photon event. 
    Energy calibration was performed by using 
    persistent environmental-background lines of $^{40}$K (1.46~MeV) and $^{208}$Tl (2.61~MeV).
    This calibration procedure for detector A was performed every 30 minutes
    to monitor and correct light yield variation of BGO which is sensitive to temperature,
    while once a day for the NaI crystal used in detector B 
    (See also ``Methods: Instrumental calibration'' in \citet{Enoto_2017} and \citet{Bowers_2017} for the calibration accuracy).

An AEF monitor, termed a field mill (Boltek EFM-100), was installed on the ground beside the building. 
    This monitor measures vertical AEF strength with a dynamic range of $\pm$5.4~kV/m at 0.5 second interval. 
    The AEF value was calibrated at the plain ground surface so that the fair-weather AEF showed around 100~V/m
    originating from the global electrical circuit.
    Because our AEF recording system lost the internet connection 
    which was used for absolute time calibration via the network time protocol during February 4 to March 21, 
    the absolute time of the AEF sampling was adjusted by the comparison between the AEF pulses and a GPS- (global positioning system) synchronized lightning catalogue
    provided by Japan Lightning Detection Network (JLDN) operated by Franklin Japan Co, ltd.

We also operated a lightning mapping system based on low-frequency (LF) radio measurements,
    hereafter the LF network, consisting of 5 stations 
    which were located in the Toyama Bay area ($\sim$60~km south from the gamma-ray observation site).
    Each station has a flat plate antenna which is sensitive to 800~Hz -- 500~kHz radio emission, 
    and its waveforms are sampled by a 4~MHz digitizer \citep{Takayanagi_2013}.
    The LF network specializes in thunderstorm observations around Toyama Bay and Noto Peninsula.
    It can determine position and timing of radio emissions 
    such as stepped leaders and main return strokes/recoil streamers of cloud-to-ground and intra/inter-cloud discharges (ICs) 
    by time-of-arrival technique.

\section{\label{sec:analysis}Analysis \& Results}
\begin{figure}[ht]
        \begin{center}
            \includegraphics[width=0.9\hsize]{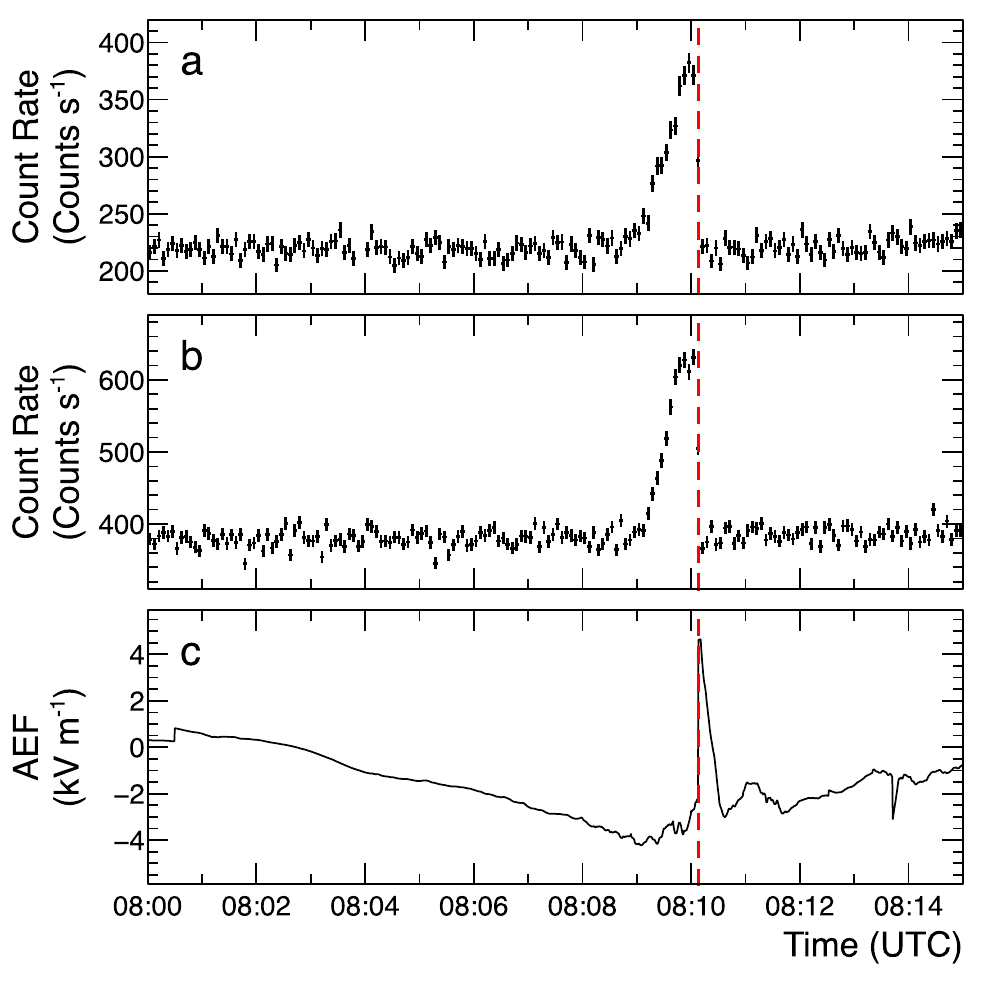}
            \caption{Histories of radiation count rates with a 5~sec binning 
            in the 0.2--7.0 MeV obtained with detector A (panel a),
            0.3--20.0 MeV with detector B (panel b),
            and calibrated AEF values (panel c) from 08:00 to 08:15 UTC on February 11, 2017.
            Negative AEF values mean upward electric field.
            Red dashed lines show time of the lightning at 08:10:08 UTC.}
            \label{fig:lightcurve}
        \end{center}
    \end{figure}

On February 11th, 2017, heavy snowfall and lightning continued along Japan Sea. 
Panels a and b of Fig.~\ref{fig:lightcurve} present count-rate histories obtained by detector A and B 
    from 08:00 to 08:15 UTC (17:00--17:15 in local time), respectively.
    Both detectors A and B recorded a count-rate increase around 08:09 
    with a time scale similar to other gamma-ray glows in winter thunderstorms \citep{Torii_2002,Tsuchiya_2007}.
    The total background-subtracted photon counts are $6640 \pm 180$ in the 0.2--7.0 MeV range for detector A, 
    and $9750 \pm 240$ in the 0.3--20.0 MeV range for detector B.
    Detection significance of this glow event is 61~$\sigma$ and 75~$\sigma$ for detector A and B respectively,
    evaluated from background fluctuation of 2-minute binned count-rate history above 3 MeV
    which energy range is not affected by washout of radioactive isotopes such as $^{214}$Bi.

The glow was then suddenly terminated and the count rate quickly returned to the background level at 08:10:08.
    Hereafter, the elapse time $t$ is defined from 08:10:08 UTC.
    The World Wide Lightning Location Network (WWLLN) reported a lightning discharge at $t=7.4$~ms.
    JLDN also reported negative and positive intra/inter-cloud discharges (ICs) around Noto peninsula at $t=7.3$~ms and $t=224.9$~ms, respectively.
    The latter occurred 2.1~km south from the observation site. Based on these measurements, 
    we consider that the sudden termination of the gamma-ray glow closely coincided with the lightning discharge.

Absolute timing of both detectors are conditioned by the GPS signals 
    and the network timing protocol service.
    Detector~A successfully received GPS signals during the observation. 
    To verify the absolute timing accuracy, laboratory experiments were performed after the observation campaign:
    pulse-per-second signals from a commercial GPS receiver were put into an analog input of detector~A, 
    and we confirmed that the timing tag of each photon is synchronized to the coordinated universal time 
    within a 5~ms systematic uncertainty.
    However, detector~B failed both to receive the GPS signal by accident and 
    to maintain the internet connection during the observation.
    Therefore, we corrected detector B timing so that time of the glow termination is consistent with that of detector~A.
    
Fitting with a Gaussian function in the range of $-300~{\rm s} < t < 60~{\rm s}$ 
    which becomes the background level in $t>0$ to imitate the sudden termination, 
    the peak time and standard deviation of the detector~A 0.2--7.0 MeV count-rate history 
    are $-5.2 \pm 3.7$~s and $29.3 \pm 2.4$~s, respectively 
    (Hereafter, all statistical errors in this paper are at 1$\sigma$ confidence level).
    In order to obtain millisecond-precision termination timing of the glow, 
    the 200~ms binning count-rate history of detector~A was fitted with a step function.
    The best-fit termination time was $t=93 \pm 52 ({\rm stat.}) \pm 5 ({\rm sys.})$~ms.

\begin{figure}[th]
        \begin{center}
            \includegraphics[width=0.9\hsize]{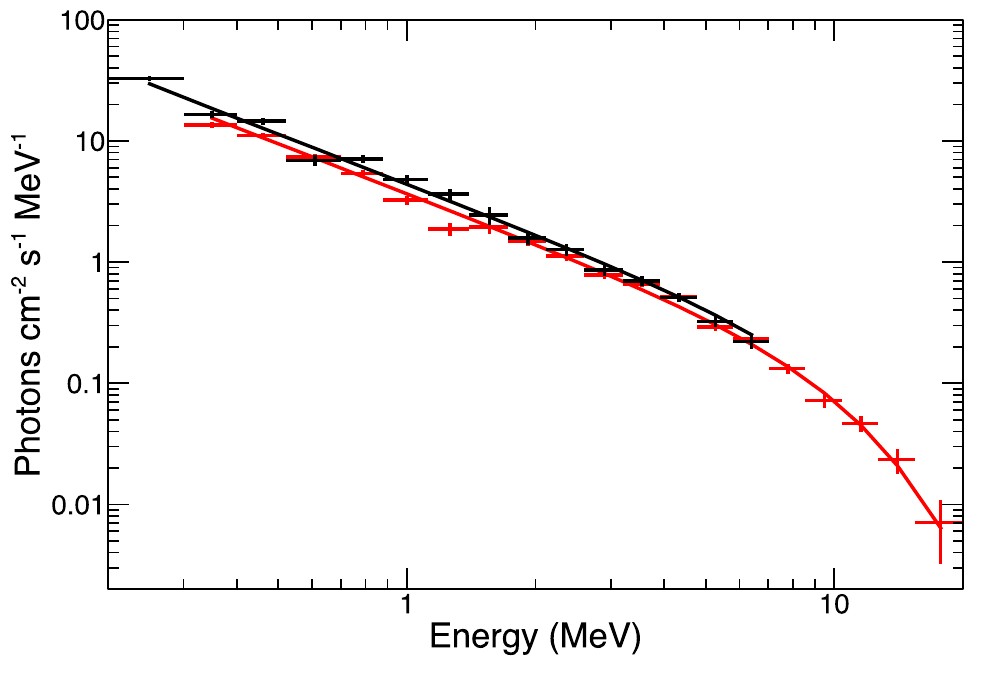}
            \caption{Time-averaged and background-subtracted 0.2--20.0~MeV spectra accumulated for $-75~{\rm s} < t < 0~{\rm s}$, 
            recorded by detector~A (black cross) and B (red) with 1~$\sigma$ statistical error bars.
            The best fit cutoff power-law model is overlaid by a solid line.
            In this panel, spectra are plotted as incident photon spectrum 
            which corresponds to the number density of gamma-ray photons 
            reached the detector, by correcting the detector energy response 
            and the effective area, while not correcting an atmospheric response.
            }
            \label{fig:spectrum}
        \end{center}
    \end{figure}

Figure~\ref{fig:spectrum} presents background-subtracted energy spectra, accumulated for $-75~{\rm s} < t < 0~{\rm s}$.
    The background is taken from $-350~{\rm s} < t < -150~{\rm s}$.
    To fit and unfold the detector-response-included spectra, we utilized a spectral analysis tool {\tt XSPEC} \citep{Arnaud_1996}, 
    which has been used for X-ray astronomy, and also available for gamma-ray spectral analysis.
    A response function of each detector was constructed by {\tt Geant4} Monte Carlo simulation \citep{Agostinelli_2003}
    and utilized as an input to {\tt XSPEC}.
    We fitted the spectra of both detectors simultaneously.
    The spectra were reproduced by a power-law function with an exponential cut-off, 
    presented as $A \times E^{-\Gamma} \exp \left[-(E/E_{\rm cut})^{\alpha} \right]$
    where $A$, $E$, $\Gamma$, $E_{\rm cut}$ and $\alpha$ are normalization 
    in units of photons~cm$^{-2}$~s$^{-1}$~MeV$^{-1}$,
    photon energy (MeV), power-law photon index, cut-off energy and cut-off index, respectively.
    We added 5\% systematic uncertainty to each bin. 
    This chi-square fitting gave a reduced chi-square 1.55 for 29 degrees of freedom, 
    which is acceptable at 5\% acceptance level. 
    The best-fit parameters are obtained as $\Gamma = 1.36 ^{+0.03}_{-0.04}$, 
    $E_{\rm cut}=11.1^{+0.8}_{-0.9}$~MeV, $\alpha = 2.0 \pm 0.3$
    and the 0.2-20.0~MeV on-ground gamma-ray photon flux $(4.14 \pm 0.14) \times 10^{-5}$~erg~s$^{-1}$~cm$^{-2}$.
    The photon index is consistent with Bremsstrahlung spectra of previous gamma-ray glows \citep{Tsuchiya_2011}.
    
The AEF history is shown in the panel c of Fig.~\ref{fig:lightcurve}.
    The AEF values were negative before, during and after the gamma-ray glow, except at the timing of the lightning.
    A steeply rising positive pulse was detected at 08:10:08.0, corresponding to the lightning discharge,
    and then the AEF exponentially decayed following the time constant of a low-pass filter embedded in the field mill.
 
\begin{figure}[t]
        \begin{center}
            \includegraphics[width=0.9\hsize]{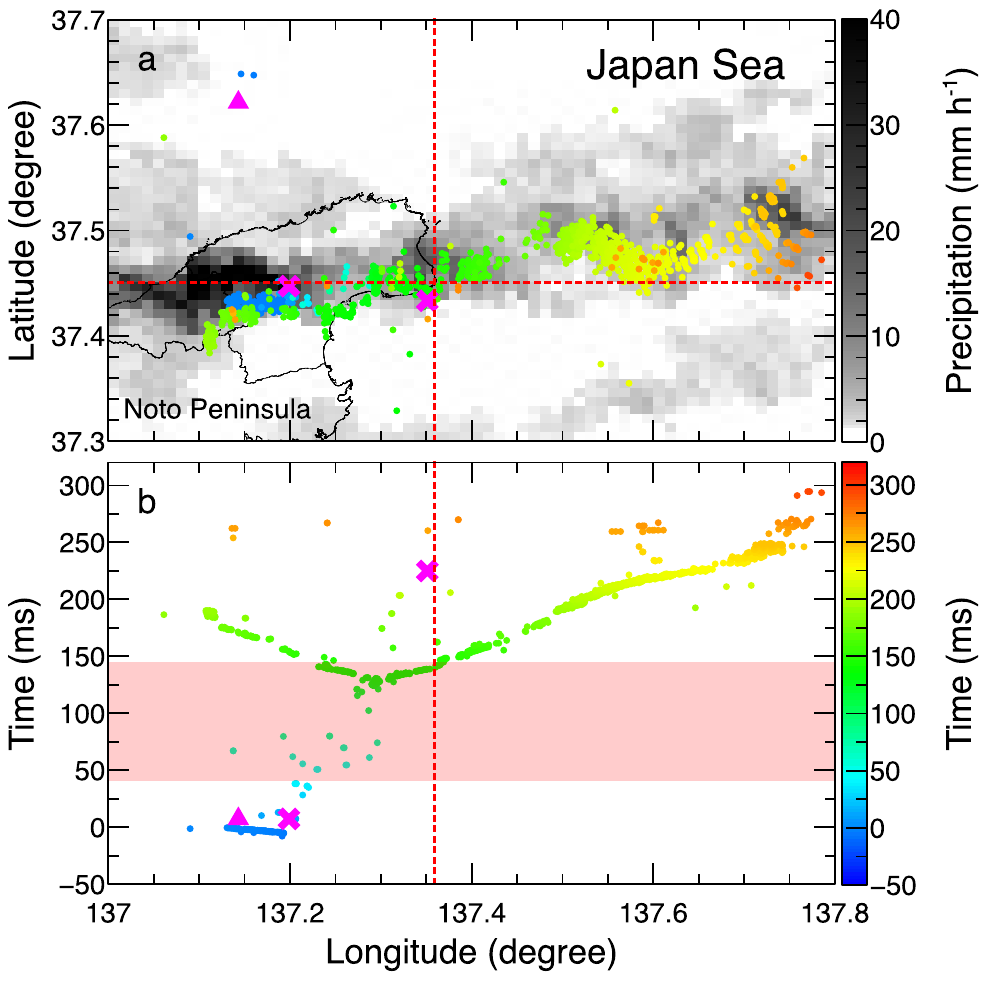}
            \caption{Time evolution of the lightning discharge (filled circle) observed in the LF band (panel a)
            and the same evolution in a form of the east-west position and time history (panel b).
            The marker color corresponds to the recorded time of discharge steps.
            The background gray-scale map is precipitation at 08:10 UTC from the C-band radar observation 
            operated by Japan Meteolorogical Agency.
            The crossed position of two red lines shows the gamma-ray observation site,
            magenta crosses the lightning positions reported by JLDN, and a magenta triangle the lightning position reported by WWLLN. 
            The red shaded region shows the moment of the gamma-ray glow termination estimated with detector~A. 
            }
            \label{fig:position}
        \end{center}
    \end{figure}

The LF network detected radio pulses at all stations at 08:10:08. 
    Figure~\ref{fig:position} presents position and time series of the LF pulse sources.
    The altitude of each LF pulse cannot be estimated because the LF stations are too far from each LF source.
    No other discharges were detected by WWLLN, JLDN, the LF network, nor the AEF monitor during the gamma-ray glow.

The LF-emitting sources spread $\sim$70~km wide in east-west direction and lasted for $\sim$300~ms.
    C-band radar operated by Japan Meteorological Agency provided a composite precipitation map at 08:10,
    shown as a gray-scale background in Fig.~\ref{fig:position}a.
    The LF emission started around 137$^{\circ}$21'~E, 37$^{\circ}$27'~N at $t \sim -10$~ms, and headed toward east, 
    which seems to have traced the intense echo area shown by the precipitation map in Fig.~\ref{fig:position}a.
    The LF sources intermittently emerged until $t \sim 120$~ms, then split into westward and eastward paths.
    The eastward path passed over Noto School.
    Such a long-distance horizontal discharge ($>$10~km) is 
    one of the still mysterious features of winter thunderstorms in the coastal area of Japan Sea \citep{Michimoto_1991,Kitagawa_1992}.
    
    \begin{figure}[]
        \begin{center}
            \includegraphics[width=0.9\hsize]{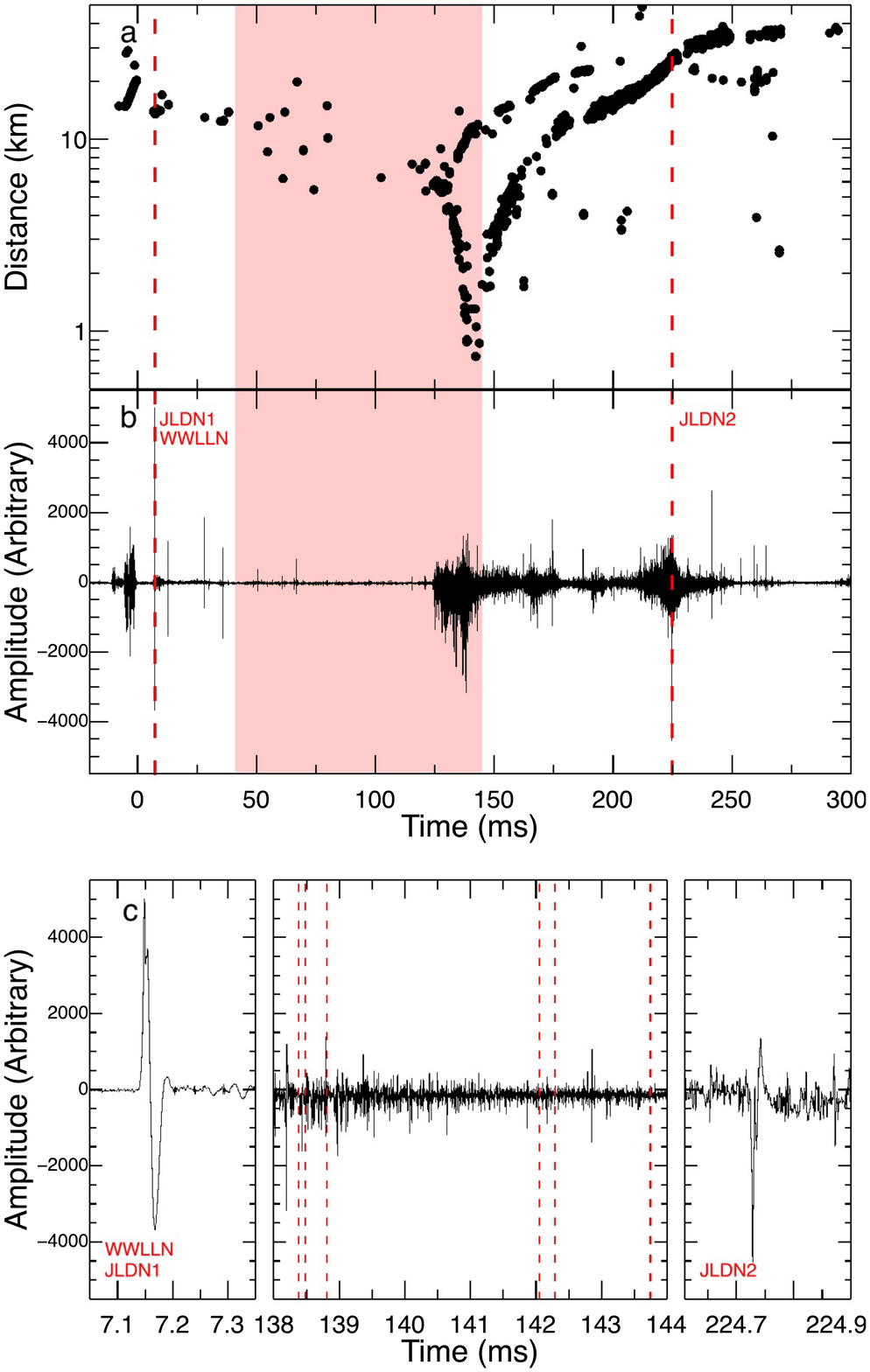}
            \caption{
            Distance from the observation site to the LF sources as a function of time, 
            converted from Fig.~\ref{fig:position} (panel a),
            LF waveforms observed at the Nyuzen station (panel b). 
            The expanded waveforms are shown in panel c 
            for JLDN/WWLLN events (left and right) and LF emissions near the observation site (center).
            Determination errors of the source position are estimated as typically 0.1~km.
            The time stamp of waveforms is uniformly shifted in order to correct propagation delay 
            between the observation site and the Nyuzen station (57~km).
            The red shaded region in panel a and b is the same as Fig.\ref{fig:position}.
            The red-dashed lines show timing of the JLDN/WWLLN events in panel b, 
            and timing of LF pulses within 1 km from the observation site in panel c center.}
            \label{fig:distance}
        \end{center}
    \end{figure}
    
Figure~\ref{fig:distance} presents horizontal distances 
    between each LF-emitting source and the gamma-ray observation site in panel a, 
    and LF waveforms observed at Nyuzen station (137$^{\circ}$30'~E, 36$^{\circ}$57'~N: 
    57~km south from the observation site) in panel b and c.
    Most of the LF pulses originate from leader development of an IC.
    Large-amplitude bipolar pulses detected at $t \sim 7$~ms and $t \sim 225$~ms 
    correspond with the negative and positive ICs reported by JLDN, respectively.
    Although the present LF observation did not allow us to determine 
    lightning types of the two large-amplitude pulses independently of the JLDN report,
    the pulses do not have strong physical connection to the present gamma-ray event 
    since they are not temporally coincident with the glow termination.
    
One of the LF sources emerged 0.7~km south-east, at the closest point, 
    from the observation site at $t=142.3$~ms.
    There are also five other sources within 1~km from the observation site.
    Therefore, the IC leader development heading eastward passed by the observation site.
    The timing of the six sources is consistent with 
    the moment of the gamma-ray glow termination within 1$\sigma$ confidence level.

There are additional information of cloud altitude.
    A ceilometer installed also on the roof of Noto School building measured 
    the cloud base altitude of 280~m at 08:09 UTC, 
    which is a typical base height for winter thunderclouds \citep{Goto_Narita_1992}.
    The low cloud base at 08:10 and heavy snowfall during 08:10-08:20 were also confirmed by
    10-minute interval images of a weather camera in Noto School.
    The C-band radar of Japan Meteorological Agency measured the radar-echo top altitude of 7~km
    at 08:10 over the gamma-ray observation site.

Short-duration gamma-ray bursts associated with lightning discharges, 
    called ``downward terrestrial gamma-ray flashes'', 
    and evidence for photonuclear reactions induced by such bursts have been detected at ground level \citep{Abbasi_2017,Bowers_2017,Enoto_2017}.
    However, neither such gamma-ray bursts nor evidence for photonuclear reactions
    at the glow termination were detected in the present case, 
    despite the IC leader development having passed near the observation site.

\section{\label{sec:discussion}Discussion}
The termination of gamma-ray glow events observed with AEF and radio observations have been already reported
    via single station measurements \citep{Chilingarian_2015,Chilingarian_2017}.
    However, in the present paper, we report the first simultaneous observation of this phenomena
    via gamma-ray, AEF, and multiple-station LF measurements which enable us to determine
    temporal and spatial evolution of discharges.

A part of LF emissions, originating from the IC leader development, 
    was detected less than 1~km away from the observation site (Fig.~\ref{fig:distance}).
    The moment of the gamma-ray termination is consistent with the time of the nearby LF emissions,
    not consistent with that of the two large-amplitude pulses.
    Therefore, it is clear that the IC leader development 
    destroyed a local structure of electric field in the thundercloud,
    which causes the termination of the gamma-ray glow, 
    despite discharge current of the IC pulses being smaller 
    than that of the large-amplitude pulses.

In the case of \citet{Tsuchiya_2013}, 
    a gamma-ray glow was terminated $\sim$800~ms before a lightning flash.
    On the other hand, JLDN detected no lightning discharges within 5~km from their observation site.
    As shown in the present study (Fig. \ref{fig:position}), the IC leader development can spread out to $\sim$70 km size.
    Among this process, the JLDN system can only detect large amplitude pulses of return strokes/recoil streamers, 
    but sometimes miss precursory discharge processes.
    Therefore, it is reasonable to assume that faint discharges
    before main discharges terminated the gamma-ray glow event in \citet{Tsuchiya_2013}.

As discussed in \citet{Chilingarian_2012}, 
    gamma-ray glows or TGEs observed by their mountain-top experiment
    are often accompanied with a lower positive charge region (LPCR) 
    which is a candidate of the electron-acceleration region, revealed from AEF measurements.
    In the present AEF observation (Fig. \ref{fig:lightcurve}c), 
    the AEF values were negative during the gamma-ray glow, 
    which basically indicates that the cloud base was negatively charged.
    On the other hand, the AEF value showed a slight positive excursion between 08:09 and 08:10.
    We can propose two possible interpretations on this result.
    
One interpretation is that an LPCR do exist, and is responsible to the electron acceleration.
    \citet{Kitagawa_1994} reported that matured winter thunderclouds have the classical tripolar charge structure 
    including an LPCR with AEF showing W-shaped temporal variation.
    While no clear W-shaped variation indicating an LPCR was found in the present data, 
    the charge structure in the thundercloud should be changed by the IC.
    Therefore, it is possible to interpret that the AEF was disturbed by the IC in the middle of W-shaped temporal variation.
    Although the AEF value was not positive before the IC, the positive excursion may originate from a weak or off-center LPCR. 
    The main electric-field provider of the gamma-ray glow should be located 
    between the LPCR and a negative charge layer above the LPCR.

The other interpretation is that the cloud base was negatively charged entirely without any LPCR structure.
    In this case, negative charge was at the bottom of the thundercloud and positive image charge was on the ground.
    A candidate of the electron acceleration site is between the negative and positive image charge layer.
    However, they were not the main electric-field provider of the gamma-ray glow
    because the AEF was not minimum when the count rate of the gamma-ray glow reached its maximum.
    A probable idea is that the charge structure for the electron acceleration 
    was located higher than the negatively-charged cloud base. 
    Namely, a local structure consisting of a negative and positive charge layers, located above the cloud base at 280~m,
    should be the main electric-field provider of the gamma-ray glow.
    Since the negatively-charged cloud base screens electric field of this local structure, 
    the structure cannot be clearly observed by AEF measurements.
    This model requires the acceleration region to be located at higher than 280~m.
    Assuming the intensity of gamma rays produced via an RREA process 
    (e.g. an typical energy distribution of terrestrial gamma-ray flashes) 
    are attenuated in the atmosphere exponentially with a folding length of 45~g~cm$^{-2}$ \citep{Smith_2010}, 
    the Bremsstrahlung gamma-ray intensity at 500~m altitude will decrease $\sim$25\% at ground level. 

\citet{Kelley_2015} also discussed, based on their airborne observation,
    that their instrument was flying at a cruise altitude of 14--15~km,
    then observed downward avalanches 
    between a main upper positive layer and a negative screening layer above.
    In our case, the cloud-top altitude was measured as 7~km.
    Assuming that avalanches were developed at higher than 5~km altitude, 
    Bremsstrahlung gamma rays can hardly reach the on-ground detectors (less than 10$^{-2}$\%).
    Therefore, this scenario is not applicable to the present glow, 
    even though another glow might have occurred at the higher region.

Based on the discussions above, we cannot conclude the charge structure 
    corresponding to the electron acceleration in the thundercloud.
    The LF network failed to evaluate the altitude of LF sources in this case,
    simply because the LF stations were located too far from the sources.
    When the source height is accurately evaluated by the LF network in similar events, 
    we will obtain an unambiguous answer to the structure of electron-acceleration regions.

In the present case, the start position of the lightning discharge 
    is 15~km west from the acceleration site (Fig. \ref{fig:position}).
    Because the start point is far from the observation site and 
    its timing is prior to the moment when the gamma-ray was terminated, 
    it is clear that the gamma-ray source did not trigger the lightning discharge.
    On the other hand, our observation directly confirmed that 
    the discharge path can pass through the charge structure emitting gamma rays.
    Continuous observations of the LF band emission and gamma-ray radiation
    are also important to reveal whether gamma-ray glows can trigger lightning discharges.

\section{\label{sec:conclusion}Conclusion}
A gamma-ray glow and its sudden termination with a lightning discharge was observed in a Japanese winter thunderstorm.
    A part of the IC leader development passing 0.7~km nearby the observation site destroyed the electron-acceleration region in a thundercloud, thus terminated the gamma-ray glow.
    The IC started $\sim$15~km far from the observation site and prior to the glow termination timing.
    Therefore, the glow did not trigger the IC in the present case.
    These results show that observations of gamma-ray glow termination events with AEF and LF lightning position measurements
    can provide clues to understand the electron-acceleration mechanisms of gamma-ray glows.

\acknowledgments
We deeply thank Shusaku Takahashi in Tokyo Gakugei University 
    and staffs of Kanazawa University Noto School
    for helping deploy and maintain the gamma-ray and AEF monitors. 
    This research is supported by the cooperative research program of Institute of Nature and Environmental Technology, 
    Kanazawa University (Accept No.16004 and No.17008),
    by JSPS/MEXT KAKENHI grant numbers 15K05115, 15H03653, 16H06006, 16K05555, 
    by Hakubi project and SPIRITS 2017 of Kyoto University,
    and by the joint research program of the Institute for Cosmic Ray Research (ICRR), the University of Tokyo.
    The GODOT deployment was supported by award AGS-1613028 from the National Science Foundation (NSF) of the United States.
    Our project is also supported by crowdfunding named Thundercloud Project,
    using the academic crowdfunding platform ``academist''.
    Y.W is supported by the Junior Research Associate program in RIKEN.
    The C-band radar data of Fig.~\ref{fig:position}a was supplied from Japan Meteorological Agency via Japan Meteorological Business Support Center.
    The geographic information in Fig.~\ref{fig:position}a was supplied from Japan Ministry of Land, Infrastructure, Transport and Tourism, and plotted by the authors.
    Our GROWTH, GODOT, AEF and LF data is available at \url{https://thdr.info/files/Wada_et_al_2018_GRL_Suzu.tar.gz}.

\listofchanges

\end{document}